\RequirePackage{fix-cm}
\documentclass[smallextended]{svjour3}       
\smartqed  
\usepackage{graphicx}
\usepackage{amsmath}
\usepackage{amssymb}
\usepackage{bbm, color}
\begin{document}

\title{Quantum simulation scheme of two-dimensional xy-model Hamiltonian with controllable coupling}



\author{Mun Dae Kim }


\institute{ \at
Korea Institute for Advanced Study, Seoul 02455, Korea \\
              \email{mdkim@kias.re.kr}           
}

\date{Received: date / Accepted: date}

\maketitle

\begin{abstract}
We study a scheme of  quantum simulator for two-dimensional
xy-model Hamiltonian. Previously the quantum simulator for a
coupled cavity array spin model has been explored, but the coupling
strength is fixed by the system parameters. In the present scheme
several cavity resonators can be coupled with each other
simultaneously via an ancilla qubit.  In the two-dimensional
Kagome lattice of the resonators the hopping of resonator photonic
modes gives rise to the tight-binding Hamiltonian which in turn
can be transformed to the quantum xy-model Hamiltonian. We
employ the transmon  as an ancilla qubit  to achieve {\it in situ}
controllable xy-coupling strength.
\end{abstract}

\section{Introduction}
In spite of the
remarkable advancements of coherent quantum operation the
realization of fully controlled quantum computing is severely
challenging in quantum information processing technology.
On the other hand, significant attention has been paid to quantum spin models as  a
promising candidate for quantum simulation of  many-body effects \cite{Georg,Cirac,Buluta}.
Quantum many-body simulation may provide a variety of possibilities to
study the properites of many-body systems, realize a new phase of
quantum matter, and eventually lead to the scalable quantum
computing, which is hard for classical approaches.

Large-scale quantum simulators
consisting of many qubits integrated have been experimentally
demonstrated to study the quantum phenomena such as many-body
dynamics and quantum phase transition. Quantum
simulators have been studied in the so-called  coupled cavity array
(CCA) model, where a two-level atom in the cavity  interacts with
its own cavity and the hopping of a photon bewteen cavities gives
rise to the cavity-cavity coupling. The CCA model has been applied
to study the Jaynes-Cummings Hubbard model (JCHM)
\cite{HartmanNP,Xue,Greentree,Schmidt,Koch,Fan}
and the Bose-Hubbard model \cite{Greentree,Koch}
to exhibit the phase transition between Mott insulator and superfluid.
However,  in the CCA model the cavity-cavity hopping amplitude is set by the
system parameters and thus not tunable.
In recent  studies for one-dimensional quantum simulators  using trapped cold atoms \cite{Nature51}
and trapped ion systems \cite{Nature53} the coupling strength was tunable.

Previously the superconducting resonators in two-dimensional lattice
have been coupled through an interface capacitance,
where the resonator-resonator coupling
strength is not controllable as the capacitance is fixed \cite{NPreview,AP}.
For superconducting resonator cavities in circuit-quantum electrodynamics (QED) systems,
qubit is located  outside of  the cavity \cite{MDK,QIP}. Hence a qubit can interact
with many resonator cavities surrounding the qubit.
By using a qubit as a mediator of coupling between many resonators
one can  obtain a tunable resonator-resonator coupling 
which is quite different from the coupling by direct
photon hopping in the CCA model.

In this study  we consider a lattice model of superconducting resonator cavities coupled
by ancilla qubits for simulating the quantum xy-model
Hamiltonian. The simulation for quantum xy-model has been studied in
one-dimensional \cite{Hartmann,Angelakis} and two-dimensional
\cite{Koch} JCHM in the CCA model architecture.
In the present model the intervening ancilla qubit which couples cavities has
controllable qubit frequency. After discarding the ancilla qubit
degrees of freedom by performing a coordinate transformation we
show that the photon states in the resonators are described by the
tight-binding Hamiltonian which, in turn, can be rewritten as the
quantum xy-type interaction Hamiltonian. Consequently, the
xy-coupling constant depends on the hopping amplitude of the
tight-binding Hamiltonian and thus on the ancilla
qubit frequency. We consider two-dimensional Kagome lattice model as well as
one-dimensional chain model for the quantum
simulation of xy-model Hamiltonian and show that the xy-coupling
strength is {\it in situ} controllable.

\section{Hamiltonian of coupled $n$-resonators}

\begin{figure}[b]
\hspace{0cm}
\vspace{0cm}
\includegraphics[width=1.0\textwidth]{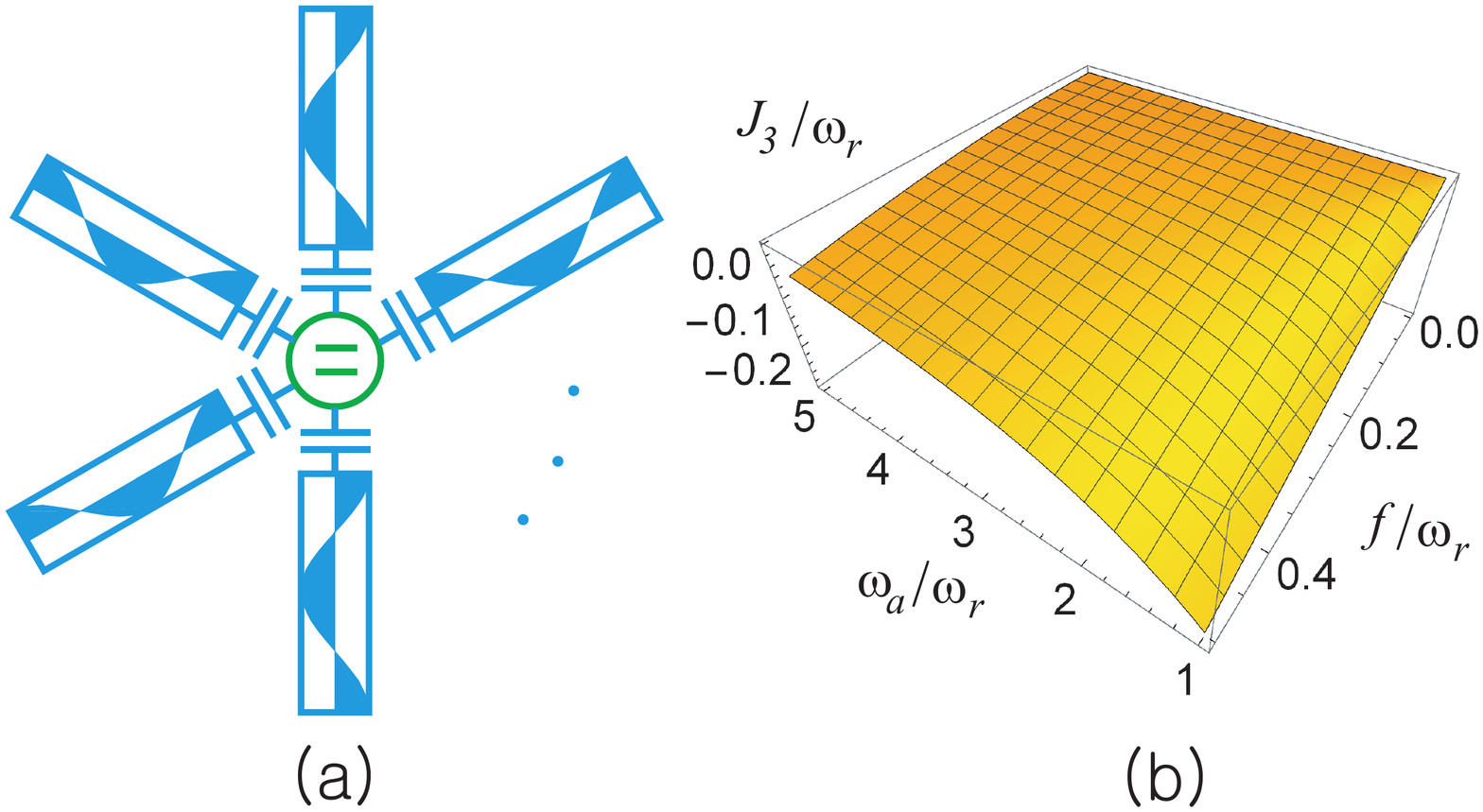}
\vspace{-2cm}
\caption{(a) $n$ cavities of circuit-QED resonators are coupled via an intervening
ancilla qubit. (b) Effective cavity-cavity coupling, $J_3$,  for $n=3$
as a function of ancilla qubit frequency $\omega_a$ and resonator-ancilla coupling
$f$ with the frequency $\omega_r$ of resonator photon mode. }
\label{fig1}
\end{figure}

In circuit-QED architectures qubits can be coupled with the transmission resonator
at the boundaries of the resonator \cite{Blais,Steffen,Inomata} so that we may couple several resonators to
a qubit as depicted in Fig. \ref{fig1} (a). In principle, any kind of qubits are available, but in this study
we employ the transmon  as the ancilla qubit coupling the resonators
with the advantage of controllability.
The Hamiltonian of the system with $n$ resonators and an ancilla qubit in Fig. \ref{fig1}(a) is given by
\begin{eqnarray}
\label{H}
H_{nR}=\frac12\omega_a\sigma^z_{a}+\sum^n_{p=1}[\omega_{rp}a^\dagger_p a_p-f_p(a^\dagger_p\sigma^-_{a}+\sigma^+_{a}a_p)],
\end{eqnarray}
where $a^\dagger_p$ and $a_p$ with the frequency $\omega_{rp}$ are the creation and annihilation operators for  microwave photon in $p$-th resonator, respectively, and the Pauli matrix $\sigma^z_{a}$ with the frequency $\omega_a$ represents the ancilla qubit state, and $f_p$ is the coupling amplitude between the photon mode in the $p$-th resonator and the ancilla qubit.
This Hamiltonian conserves the excitation number
\begin{eqnarray}
\label{Ne}
{\cal N}_e=\sum^n_{p=1}N_{rp}+(s_{az}+1/2),
\end{eqnarray}
where $s_{az}\in \{-1/2,1/2\}$ are the eigenvalue of the operator $S_{az}=\frac12\sigma^z_{a}$
for ancilla qubit and $N_{rp}$ is the excitation number of oscillating mode in $p$-th resonator.
Here, we  consider the subspace that ${\cal N}_e=1$ and thus
$N_{rp}\in \{0,1\}$, that is, the state of resonator is the
superposition of zero and one-photon states which was
generated in experiments previously \cite{Houck,Hofheinz08,Hofheinz09}.

In order to obtain the Hamiltonian describing the interaction between
photon modes  we introduce the transformation
\begin{eqnarray}
\label{tildeH}
{\tilde H}_{nR}=U^\dagger H_{nR} U,
\end{eqnarray}
where
\begin{eqnarray}
U=e^{-\sum^n_{p=1}\theta_p(a^\dagger_p\sigma^-_{a}-\sigma^+_{a}a_p)}.
\end{eqnarray}
Here we, for simplicity, assume identical resonators and thus set  $\omega_{rp}=\omega_r,
f_p=f$ and $\theta_p=\theta$.
We then expand $U=e^M$ with $M=-\sum^n_{p=1}\theta_p(a^\dagger_p\sigma^1_{a}-\sigma^+_{a}a_p)$
by using the relation $e^M=1+M+\frac{1}{2!}M^2+\frac{1}{3!}M^3+\cdots$ to obtain
\begin{eqnarray}
U_{pp}\!\!&=&\!\!1-\frac{1}{2!}\theta^2\!\!+\!\!\frac{1}{4!}n\theta^4\!\!-\!\!\frac{1}{6!}n^2\theta^6
\!\!+\!\!\cdots= \frac{1}{n}(n\!\!-\!\!1\!\!+\!\!\cos\sqrt{n}\theta)\\
U_{n+1,n+1}\!\!&=&\!\!1-\frac{1}{2!}n\theta^2+\frac{1}{4!}n^2\theta^4-\frac{1}{6!}n^3\theta^6+\cdots= \cos\sqrt{n}\theta\\
U_{p,n+1}\!\!&=&\!\!-\theta\!\!+\!\!\frac{1}{3!}n\theta^3\!\!-\!\!\frac{1}{5!}n^2\theta^5\!\!+\!\!\cdots
= -\frac1{\sqrt{n}}\sin\sqrt{n}\theta=-U_{n+1,p}\\
U_{pq,p\neq q}\!\!&=&\!\!-\frac{1}{2!}\theta^2\!\!+\!\!\frac{1}{4!}n\theta^4\!\!-\!\!\frac{1}{6!}n^2\theta^6\!\!+\!\!\cdots=
\frac{1}{n}(\cos\sqrt{n}\theta\!\!-\!\!1).
\end{eqnarray}
Here $U$ is a $(n+1)\times (n+1)$ matrix in the basis of
$|N_{r1},N_{r2},N_{r3}, \cdots, N_{rn},s_{az}\rangle$ and
$p,q \in \{1,2,3, \cdots, n\}$.

The degree of freedoms of ancilla qubit and resonator photon modes
in the Hamiltonian of Eq. (\ref{H}) can be decoupled by imposing the
condition
\begin{eqnarray}
\tan2\sqrt{n}\theta=2\sqrt{n}\frac{f}{\Delta}
\end{eqnarray}
which can be achieved by adjusting the detuning  $\Delta\equiv \omega_a-\omega_r$ \cite{Blais}.
The resulting transformed Hamiltonian of Eq. (\ref{tildeH}) becomes
\begin{eqnarray}
\label{tildeHM} {\tilde H}_{nR}=\left(
\begin{array}{cccccc}
\epsilon^r_1 & J_n & J_n & \cdots & J_n &0 \\
J_n & \epsilon^r_2 & J_n & \cdots & J_n &0 \\
J_n &  J_n & \epsilon^r_3 & \cdots & J_n &0 \\
\vdots & \vdots & \vdots & \ddots   &\vdots  &\vdots \\
J_n &  J_n  & J_n &\cdots &  \epsilon^r_n &0 \\
0 & 0 & 0& \cdots & 0 & \epsilon^a
\end{array}
\right),
\end{eqnarray}
where  $\epsilon^a$ is the energy for the state that  $s_{az}=1/2$
and $N_{rp}=0$ for all $p \in \{1,2,3,\cdots,n\}$, and
$\epsilon^r_p$ is the energy for the state that $s_{az}=-1/2$ and only the $p$-th resonator has
one photon, $N_{rp}=1$ and $N_{rq}=0 ~(q\neq p)$.
For identical resonators, $\epsilon^r_1=\epsilon^r_2=\epsilon^r_3= \cdots=\epsilon^r$
and $\epsilon^a$ are explicitly evaluated as
\begin{eqnarray}
\epsilon^r&=&-\frac{1}{2n}\left(\Delta+ sgn(\Delta)\sqrt{\Delta^2+4nf^2}\right)+\frac12\omega_r,\\
\epsilon^a&=&\frac12
sgn(\Delta)\sqrt{\Delta^2+4nf^2}+\frac12\omega_r,
\end{eqnarray}
and the resonator-resonator coupling is given by
\begin{eqnarray}
\label{J}
J_n=\frac{1}{2n}\left(\Delta-sgn(\Delta)\sqrt{\Delta^2+4nf^2}\right),
\end{eqnarray}
where  $sgn(\Delta)$ is $+1(-1)$ for $\Delta>0~(\Delta<0)$.

In the subspace satisfying   ${\cal N}_e=1$
the Hamiltonian ${\tilde H}_{nR}$ in Eq. (\ref{tildeHM})
can be represented as
\begin{eqnarray}
\label{TB}
{\tilde H}_{nR}&=&\frac12\sum^2_{p=1}\omega'_r(2a^\dagger_pa_p-1)
+\sum^n_{p,q=1,p\neq
q}J_n(a^\dagger_pa_q+a_pa^\dagger_q)\nonumber\\
&&+\frac12\omega'_a\sigma^z_{a}.
\end{eqnarray}
Consequently, $\epsilon^r_p$ and $\epsilon^a$ can be rewritten as $\epsilon^r_p=\epsilon^r=-\frac{n-2}{2}\omega'_r-\frac12\omega'_a$
and  $\epsilon^a=-\frac{n}{2}\omega'_r+\frac12\omega'_a$
so that we can have the relations, $\omega'_a=-(n\epsilon^r-(n-2)\epsilon^a)/(n-1)$ and
$\omega'_r=-(\epsilon^r+\epsilon^a)/(n-1)$.
In this tight-binding Hamiltonian the ancilla qubit operator $\sigma^z_{a}$ is decoupled from the
resonator photon mode $a$, and afterward we will ignore the ancilla term.

The tight-binding Hamiltonian ${\tilde H}_{nR}$  can be easily transformed to the xy-spin model by
introducing a pseudo spin operator $\sigma_p$ such that
$2a_pa^\dagger_p-1=|1_p\rangle\langle 1_p|-|0_p\rangle\langle 0_p|=\sigma^z_{p}$ and
$a^\dagger_pa_q+a_pa^\dagger_q=|1_p0_q\rangle\langle 0_p1_q|+|0_p1_q\rangle\langle 1_p0_q|
=\sigma^+_{p}\sigma^-_{q}+\sigma^-_{p}\sigma^+_{q}=(1/2)(\sigma^x_{p}\sigma^x_{q}+\sigma^y_{p}\sigma^y_{q})$
as follows:
\begin{eqnarray}
\label{xy}
H_{xy}=\frac12\sum^n_{p=1}\omega'_r\sigma^z_{p}
+\frac12\sum^n_{p,q=1,p\neq
q}J_n(\sigma^x_{p}\sigma^x_{q}+\sigma^y_{p}\sigma^y_{q}).
\end{eqnarray}
Here, the hopping parameter $J_n$ acts as a xy coupling constant
between pseudo spins.

\section{xy-model with tunable coupling  }

\begin{figure}[b]
\hspace{2cm}
\includegraphics[width=0.7\textwidth]{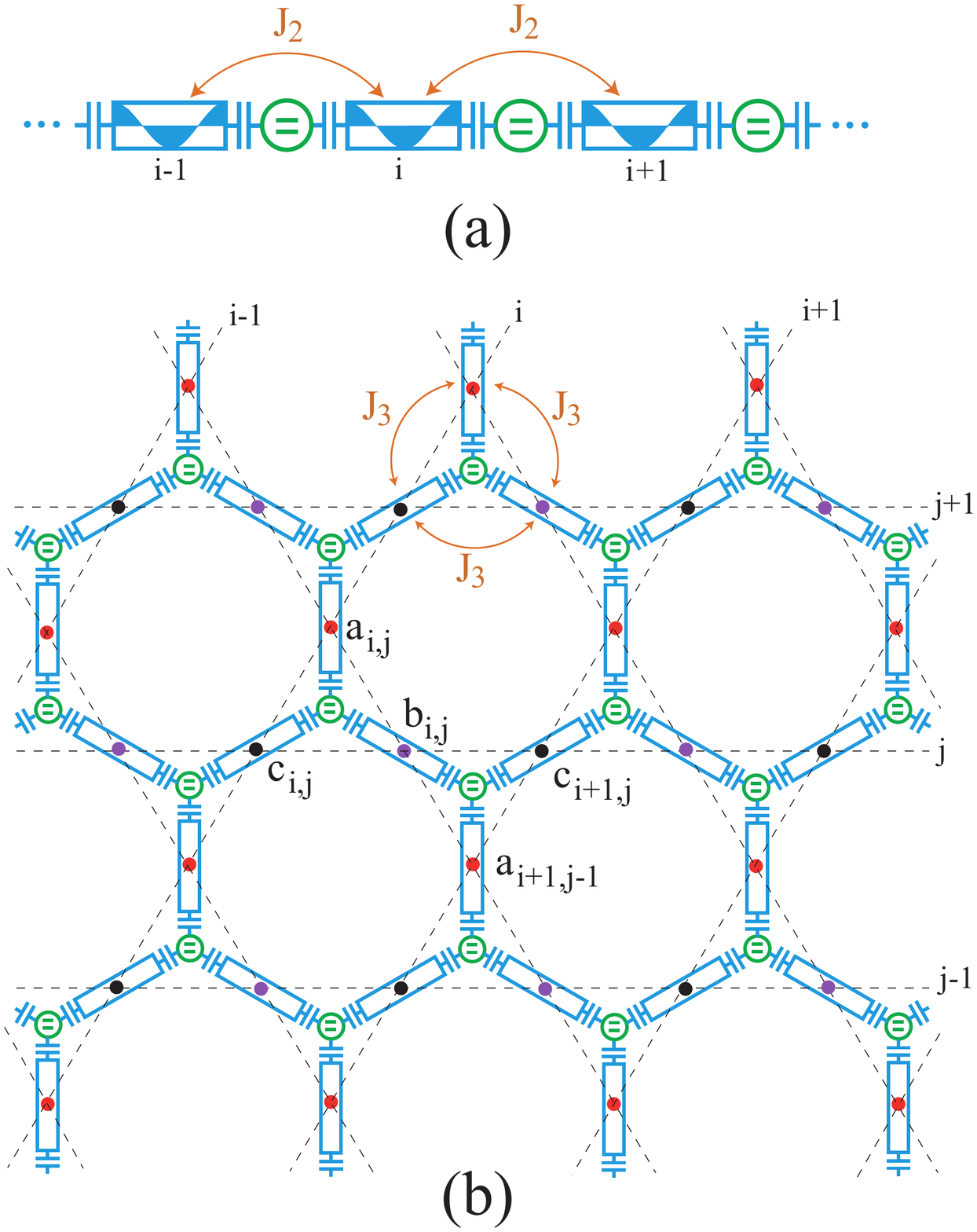}
\vspace{-1cm}
\caption{(a) One-dimensional chain of cavity
resonators coupled via ancilla qubits with the
effective cavity-cavity coupling $J_2$. (b) Two-dimensional Kagome
lattice of cavity resonators consisting of three triangular
sublattices, $a_{i,j}$ (red), $b_{i,j}$(purple) and
$c_{i,j}$(black), with effective coupling strength $J_3$.}
\label{fig2}
\end{figure}

Figure \ref{fig2}(a) shows one-dimensional lattice model
by extending the structure in Fig. \ref{fig1}(a) for two resonators and an ancilla qubit ($n=2$).
The transformation of Hamltonian ${\tilde
H}_{2R}=U^\dagger H_{2R} U$ in Eq. (\ref{tildeH}) can be evaluated
 by using the transformation matrix
$U=e^M=e^{-\sum^2_{j=1}\theta_j(a^\dagger_j\sigma^--\sigma^+a_j)}$
with
\begin{eqnarray}
H_{2R}=\left[ {\begin{array}{ccc}
   \omega_{r1}-\frac12\omega_a  & 0 & -f_1\\
   0 & \omega_{r2}-\frac12\omega_a  & -f_2 \\
   -f_1  & -f_2 & \frac12\omega_a \\
  \end{array} } \right],
M=  \left[ {\begin{array}{ccc}
   0  & 0 & -\theta_1\\
   0 & 0  & -\theta_2 \\
   \theta_1  & \theta_2 & 0 \\
  \end{array} } \right].
\end{eqnarray}
For identical resonators such that
$\omega_{r1}=\omega_{r2}=\omega_r, f_1=f_2=f$, and thus
$\theta_1=\theta_2=\theta$, the transformation matrix can be
calculated as
\begin{eqnarray}
U= \left[ {\begin{array}{ccc}
   \frac12(\cos\sqrt{2}\theta+1)& \frac12(\cos\sqrt{2}\theta-1) & -\frac1{\sqrt{2}}\sin\sqrt{2}\theta  \\
     \frac12(\cos\sqrt{2}\theta-1) & \frac12(\cos\sqrt{2}\theta+1) & -\frac1{\sqrt{2}}\sin\sqrt{2}\theta  \\
   \frac1{\sqrt{2}}\sin\sqrt{2}\theta  & \frac1{\sqrt{2}}\sin\sqrt{2}\theta & \cos\sqrt{2}\theta  \\
  \end{array} } \right]
\end{eqnarray}
with the basis $|N_{r1},N_{r2},s_{az}\rangle \in
\{|1,0,-1/2\rangle,|0,1,-1/2\rangle,|0,0,1/2\rangle\}$, the
photon number in 1st (2nd) resonator $N_{r1} (N_{r2})$ and the
ancilla qubit spin  $s_{az}$.

The transformed Hamiltonian ${\tilde H}_{2R}$  can be represented as the tight-binding
Hamiltonian of Eq. (\ref{TB}),
\begin{eqnarray}
{\tilde H}_{2R}=\frac12\sum^2_{i=1}\omega'_r(2a^\dagger_ia_i-1)
+\sum^N_{i=1}J_2(a^\dagger_ia_{i+1}+a^\dagger_{i+1}a_i),
\end{eqnarray}
with the hopping parameter
$J_2=\frac14(\Delta-\sqrt{\Delta^2+8f^2})$, discarding the decoupled ancilla term.
This tight-binding Hamiltonian describes photon hopping in the chain model of Fig. \ref{fig2}(a),
which can be subsequently transformed to the
one-dimensional xy-model Hamiltonian similar to Eq. (\ref{xy}) as
\begin{eqnarray}
H^{1D}_{xy}=\frac12\sum^N_{i=1}\omega'_r\sigma^z_{i}
+\frac12\sum^N_{i=1}J_2(\sigma^x_{i}\sigma^x_{i+1}+\sigma^y_{i}\sigma^y_{i+1}).
\end{eqnarray}

Further, for $n=3$  we can construct a two-dimensional lattice model as shown in Fig. \ref{fig2}(b).
Here the ancilla qubits form the hexagonal lattice, but the resonators the dual lattice, i.e., the Kagoma lattice.
The Kagome lattice has been widely  studied in the relation of, for example,
the frustrated spin model \cite{Mielke} and the interacting boson model \cite{You,Petrescu}.
The Kagome lattice in  Fig. \ref{fig2}(b) consists of three triangular sublattices
denoted as $a_{i,j}, b_{i,j}$ and $c_{i,j}$.
Here, two triangles consisting of, for example, $a_{i,j}, b_{i,j}, c_{i,j}, a_{i+1,j-1}$
and $c_{i+1,j}$ in Fig. \ref{fig2}(b),
make up the unit cell and thus the  xy-model Hamiltonian
in the Kagome lattice can be written as
\begin{eqnarray}
\label{Kagome}
H^{Kagome}_{xy}&=&\frac12\sum^N_{i,j=1}\omega'_r(\sigma^z_{a,i,j}+\sigma^z_{b,i,j}+\sigma^z_{c,i,j})\nonumber\\
&&+\frac12\sum^N_{i,j=1}J_3(\sigma^x_{a,i,j}\sigma^x_{b,i,j}+\sigma^x_{b,i,j}\sigma^x_{c,i,j}+\sigma^x_{c,i,j}\sigma^x_{a,i,j}\nonumber\\
&&+\sigma^y_{a,i,j}\sigma^y_{b,i,j}+\sigma^y_{b,i,j}\sigma^y_{c,i,j}+\sigma^y_{c,i,j}\sigma^y_{a,i,j}\nonumber\\
&&+\sigma^x_{a,i+1,j-1}\sigma^x_{c,i+1,j}+\sigma^x_{b,i,j}\sigma^x_{a,i+1,j-1}+\sigma^x_{c,i+1,j}\sigma^x_{b,i,j}\nonumber\\
&&+\sigma^y_{a,i+1,j-1}\sigma^y_{c,i+1,j}+\sigma^y_{b,i,j}\sigma^y_{a,i+1,j-1}+\sigma^y_{c,i+1,j}\sigma^y_{b,i,j}).\nonumber\\
\end{eqnarray}

Photons hop between resonators with amplitude $J_n$  which depends
on the sign of detuning $\Delta$ in Eq. (\ref{J}). If
$\Delta>0$, the hopping amplitude is negative, $J_n<0$, indicating
that the hopping process reduces the total system energy and the
photons hop between cavities, while for $\Delta<0$ and  $J_n>0$
the hopping process has energy cost and thus the photon state is
localized in the resonator at the ground state. Since typically
the transmon qubit frequency $\omega_a/2\pi \sim$ 10GHz
\cite{KochPRA,Wallraff} and the resonator microwave photon
frequency in circuit-QED scheme is $\omega_r/2\pi\sim$ 5-10GHz
\cite{Blais}, we will consider the  parameter range of
$\Delta=\omega_a-\omega_r>0$.

For three resonators coupled to an ancilla qubit ($n=3$) in Fig.
\ref{fig1}(a) the hopping amplitude becomes
$J_3=\frac{1}{6}(\Delta-\sqrt{\Delta^2+12f^2})$. Figure
\ref{fig1}(b)  shows $J_3$ as a function of the ancilla qubit
frequency $\omega_a$ and the ancilla-resonator coupling strength
$f$. For the resonant case, $\Delta=\omega_a-\omega_r=0$, the
hopping ampltude has the maximum value, $|J_3|=f/\sqrt{3}$, and
diminishes as the detuning $\Delta$ grows, which means that $J_3$
can be controllable between $-f/\sqrt{3}<J_3 <0$. Here
the typical value of the coupling between transmon ancilla and resonator
$f/2\pi\sim$ 100MHz \cite{Zeytin,Keller,Bosman}.

If we can adjust the parameters, $\Delta=\omega_a-\omega_r$ and
$f$, the coupling constant $J_3$ becomes tunable.  The resonator
frequency $\omega_r$ and the resonator-photon coupling $f$ are
usually set in the experiment, but we can tune the ancilla qubit frequency
$\omega_a$ during the experiment  for some qubit scheme. For the
transmon qubit the qubit frequency is represented as $\omega_a\sim
\sqrt{8E_JE_C}$ with the Josephson coupling energy $E_J$ and the
charging energy $E_C$ \cite{KochPRA}. Since the Josepson coupling
energy $E_J=E_{J,max}|\cos(\pi\Phi/\Phi_0)|$ is controllable by
varying the magnetic flux $\Phi$ threading a dc-SQUID loop
\cite{KochPRA}, we can adjust the frequency of the transmon qubit, $\omega_a$.
In the Hamiltonian for the two-dimensional xy-model in Kagome lattice in  Eq. (\ref{Kagome})
$J_3$, corresponding to the coupling constant between pseudo spins $\sigma$,
becomes tunable.
Hence, in this way we can achieve a quantum simulator for the
two-dimensional xy-model in Kagome lattice with {\it in situ}
tunable coupling.

We can measure the resonator states by attaching
measurement ports to the resonators, resulting in a complex lattice design.
Instead, as in a recent study \cite{Kollar} measurement ports can be attached
at the boundary of the lattice, but the analysis of the simulation results becomes complicated.
In this study we assume identical resonators with equal ancilla qubit-resonator coupling $f$
and further consider  a restricted subspace with ${\cal N}_e=1$ in the Hilbert space as shown
in Eq. (\ref{Ne}).  If the couplings $f_p$ have  some fluctuations from the uniform value $f$,
the transformed Hamiltonian will deviate from  the exact xy-model Hamiltonian.
Furthermore, multiple photons or higher harmonic modes in the resonators  may be generated,
giving rise to errors in the processes. The effect of these  non-idealities should be considered  in a future study.

\section{conclusion}

We proposed a scheme for simulating quantum xy-model Hamiltonian
in two-dimensional Kagome lattice of resonator cavities with
tunable coupling. By using an intervening ancilla qubit several
cavities are coupled with each other. We found that the cavity
lattice formed by extending this structure can be transformed to
the tight-binding lattice of photons after discarding the ancilla qubit
degree of freedom. In the subspace of zero and one photon mode in
the cavities this Hamiltonian can be described as the quantum
xy-model Hamiltonian. We introduced the ancilla transmon qubit
whose energy levels can be controlled by varying a threading
magnetic flux. The coupling strength can be {\it in situ} tuned by
adjusting the frequency of ancilla qubit intervening cavities.

\begin{acknowledgements}
This work was  supported  by the Basic
Science Research Program through the National Research Foundation
of Korea (NRF) funded by the Ministry of Education, Science and
Technology (2011-0023467).
\end{acknowledgements}





\end{document}